\documentstyle[aps,twocolumn]{revtex}
\begin{document}
\title{Ab-initio study of oxygen vacancies in 
$\alpha$-quartz}
\author{C. M. Carbonaro, Vincenzo Fiorentini, and S. Massidda }
\address{Istituto Nazionale di Fisica della Materia  and
 Dipartimento di Scienze Fisiche,\\ Universit\`a di Cagliari, 
via Ospedale 72, I-09124 Cagliari, Italy\\}
\maketitle
\draft
\begin{abstract}
Extrinsic levels, formation energies, and
relaxation geometries are calculated ab initio for oxygen vacancies in
$\alpha$-quartz SiO$_2$. 
The vacancy is found to be thermodynamically
stable in the charge states $Q$=+3, $Q$=0,
$Q=$--2, and $Q$=--3. The  charged states
are stabilized by
large and asymmetric distortions near the vacancy site.  Concurrently, 
Franck-Condon shifts for absorption and recombination related 
to these states are found to be strongly asymmetric.
In undoped quartz,  the ground state of the vacancy is the neutral
charge state,   
while for moderate $p$-type and $n$-type doping, the +3 and   --3
states are  favored, respectively, over  a wide Fermi level window.   
Optical transitions related to the vacancy are predicted at  
around 3 eV and
6.5 eV (absorption) and 2.5 to 3.0 eV (emission),  depending 
 on the charge state of the ground state.
\end{abstract}

\section{Introduction}
In this paper we report  {\it ab initio} calculations of
equilibrium geometries, formation energies, and extrinsic gap levels
of  vacancies in $\alpha$-quartz.  We concentrate 
on oxygen vacancies, which are common defects in both silica
(amorphous SiO$_2$) and quartz, and alleged to be precursors
of other
more complex defects \cite{add1-complexdef}. 
In view of   the short-range 
order conservation upon thermal amorphization of quartz, which 
basically rules out
complex multicenter native defects,
the state-of-the-art results for oxygen 
vacancies in crystalline quartz to be discussed below are
expected to shed some light on defect phenomenology in
amorphous silica. Recent ab initio molecular
dynamics simulations of the amorphization of a
small (128-atom) sample of $\alpha$-quartz via melt
quenching \cite{pasq} have given evidence for an almost 
perfect conservation of short range order in the amorphous phase; the
structure obtained in the  simulation \cite{pasq} maintains 
chemical order, with Si-centered tetrahedra linked by corner-sharing
oxygens.  In these simulations, disorder manifests itself 
merely as a broadening of the
distribution of Si-O-Si angle between linked tetrahedra; the O-Si-O
angle distribution has instead a sharp maximum close to the
tetrahedral angle 
of 109$^{\circ}$.  Edge-sharing tetrahedra are absent in the amorphous
phase, but they occasionally appear in the liquid \cite{pasq}. On the other
hand, even in the liquid the dominant defects are simple miscoordinated
 atoms (3- or 5-fold coordinated Si, and 1- or 2-fold coordinated O),
occurring with small ($\sim$ 10 \%) probability. 
Although  the  predictions of ab initio 
molecular dynamics may be considered 
too optimistic (mostly, because of the smallness of the simulated 
amorphous sample), it is clear that even the simplest defects are not 
easily formed thermally in SiO$_2$. More complex defects are 
 likely to be even more rare. A study of lattice vacancies seems
therefore timely, also because, although many semiempirical studies
\cite{add4-fowler} have appeared in the past,  even
 these simple defects  have  not been widely studied at
the ab initio level (recent density-functional-theory studies
are reported in \cite{add2-allan} and \cite{add3-pasq}). In particular,
no systematic study exists, to our knowledge, dealing with the 
charge states of the O vacancy beyond the neutral and +1 states 
\cite{add2-allan,add3-pasq}.

\section{Method}

We perform  ab initio
density-functional theory (DFT)\cite{dft} plane-waves pseudopotential
calculations, using  ultrasoft-pseudopotentials \cite{vand}, which
guarantee a high degree of transferability and a fast convergence in
Fourier space of all the relevant quantities \cite{vand}.  A plane-waves 
basis is used with a cut-off at 20 Ryd. A conjugate-gradient total-energy
minimization \cite{ksv} is used to obtain the self-consistent electronic 
ground-state at fixed ions, and an efficient
relaxation method \cite{bern} is used to reach the minimum energy structure
following Hellmann-Feynman forces.  We employ 36-atom orthorombic
supercells with no imposed symmetry, and one special {\bf k} point as
Brillouin zone sampling.  This setting gives structural properties in
 agreement with those obtained by Liu {\it et al.}
\cite{garof}, matching experiment to better than 1 \%, apart from two
O internal parameters having a deviation of $\sim$ 2\%.

The formation energy for an O vacancy in SiO$_2$ 
 in charge state {\it Q}
 (where, conventionally, $Q$ is the
number of electrons transferred  from the defect  to a
reservoir of chemical potential $\mu_e$)  is given by 
\begin{equation}
E_f(Q) =  \Delta E^{\rm tot}(Q)   - \mu_{\rm O} + Q (\mu_e + E_v^Q),
\label{eform}
\end{equation}
where
$\Delta E^{\rm tot}(Q)$ is the total energy difference between
the  defected supercell in 
 charge state $Q$ and the perfect bulk supercell, and
 $\mu_e$ is the electron chemical potential
(i.e the Fermi level in our T=0 calculation). The  zero 
of $\mu_e$ may be assumed to be 
 the top valence band energy
$E_v^Q$ of the defected crystal in the charge state $Q$. 
The chemical potential $\mu_{\rm O}$
is one half of the total energy of the 
spin-unpolarized O$_2$ molecule (calculated in a large supercell).
As expressed  above, the formation energy depends
on the charge state and on the Fermi level. The latter
embodies the effects of background 
doping, competing defect species, or externally injected carriers.

\subsection{Defect levels as total energy differences}

The eigenvalues of the hamiltonian
of a  crystal with a  fixed number of electrons
have no rigorous relationship to observable optical
transition, unless they are many-body quasiparticle energies.
This is also the case for defected crystals, and for 
transitions involving  extrinsic levels therein.
In particular, transitions to and from extrinsic levels 
cannot simply be extracted  as differences of DFT
 eigenvalues of e.g. the neutral state
of the defect. 
This is true for any one-body theory, including
Hartree-Fock calculations \cite{add6-snyder}, since  Koopman's theorem
gives (e.g.) an  ionization potential which is not exact even in principle.
Within DFT one can resort to total energy differences, since
the DFT total energy is exact in principle,
 apart from the local density approximation (LDA)  and other technical
issues of practical relevance \cite{add7-pickett}.
 In particular, in a DFT calculation 
one calculates   formation (i.e. total)  energy differences
 between the relevant systems. For e.g. the case of $Q$=0 and $Q$=--1,
this difference is,  by definition, the energy needed to extract an electron
from the $Q$=--1 state (ionization of the charged center) leaving
behind a $Q$=0 center, 
or to add one to the $Q$=0 state (binding  to the neutral center)
transforming it into a $Q$=--1 state. When account for geometrical
changes upon occupation  is taken, this difference is the thermal excitation 
level; ``thermal'' indicates that  the structure is allowed sufficient time
to rearrange to its new equilibrium position. We concentrate on the 
calculation of this quantity, and discuss later optical transitions.

Apart from basic choices such as the use of the LDA as opposed to 
gradient-corrected functionals, the uncertainties in the predicted 
defect energy levels arise mainly from the alignment of the energy zero of 
 differently-charged supercells. Since we choose 
conventionally the Fermi level zero at  the valence band top,  $Q\neq$ 0
charge states have a constant energy offset $-Q E_v^Q$ away from the 
neutral state energy (Eq. \ref{eform}).  We proceed by calculating 
the {\it  bulk} valence band top $E_v$  in a
bulk supercell, and express  the valence 
top in the charged supercell as
\begin{equation} E_v^Q = E_v + \Delta V,\end{equation}
where  $\Delta V$ is the difference of
the macroscopic averages\cite{bald} of the local potentials of
the  charged  supercell and of the  bulk supercell,
\begin{equation}\Delta V =  \overline{V}_{\rm Q}
-  \overline{V}_{\rm bulk}.\end{equation} This quantity
is evaluated  in an appropriate bulk-like region, i.e. 
regions of the simulation cell where the local
 potential is  unaffected by the defect. 
The resulting uncertainties are at most of order  $\pm$ 0.1 eV in the
thermal ionization levels.

\section{Results}

For the neutral state of the vacancy, 
 the formation energy is 6.97  eV. Such large energy 
is in accord with the need for irradiation 
 to produce vacancies in the first place.
 As seen from Fig. \ref{f1eform}, 
the vacancy is  stable in the $Q$=+3 state for   Fermi level values of
up to  2 eV above the valence band top.
Thereafter, the neutral state becomes stable, and remains such 
for Fermi levels up to about  5.5 eV. 
The neutral state  is therefore the ground state of the oxygen 
vacancy in nominally undoped quartz, i.e. for the Fermi level at 
mid-gap ($\sim$ 4.5 eV).
Increasing further the Fermi level, the vacancy becomes negatively
 charged. 
The $Q=-3$ charge state is the most stable state for Fermi levels
 ranging from about 5.5 eV to 8 eV (neglecting  the 
small Fermi level range from 5.5 eV to 5.7 eV,   where the --2 state 
prevails). 

While the thermal level  sequence  which is usually  expected is 
$Q$ = +3, +2, +1,  0, --1, --2, --3 ...,
in the present case we find  a negative-U behavior \cite{add8-scherz}
between the triply
positive state
and the neutral state, and between the neutral 
and the $Q=-3$ state:
 the $Q=$+2 and  $Q=$+1 states are unstable   
against the release of a further electron, and 
 the $Q$=--1
and  $Q$=--2 states are unstable towards
the capture of a further electron. This is due to a strong electron-lattice
coupling, as is discussed below. 
The realization of large-$Q$ negative and positive  states is 
not unusual for wide gap materials.

\subsection{Defect electronic structure}

The $Q=0$ vacancy is a  symmetric combination of 
Si dangling bonds. The Si neighboring to the  vacancy relax 
almost symmetrically, by 9.5 \% of the unrelaxed Si-O bond length,
towards the vacant site. In Fig. \ref{f2rho0},  we display 
density isosurfaces for the defect state of the neutral vacancy.

The only stable positive charge state is the $Q$=+3
state. The two Si atoms neighboring to the vacancy 
relax strongly away from
their original positions in an asymmetric fashion: 
their distance to the vacancy site increases
from  1.62 \AA\, to 2.16 \AA, and from 1.61 \AA\ to 2.81 \AA, 
respectively. The two Si atom are three-fold coordinated and close to
co-planar with the neighboring oxygens. The charge density of the last
occupied state is delocalized, suggesting that the   defect level is
resonant with the valence  band.
The thermodynamical  instability of the singly  positive  vacancy
signals potential difficulties in previously proposed 
models \cite{add4-fowler} of the so called  E$'$ center.

The $Q=-1$ charge state is the antibonding combination of the same
 dangling bonds as above, as can be seen in Fig. \ref{f3rho-1}
(compare  with the previous Figure).
Additional relaxations with respect to
the neutral state are below 1\%. Thus, electronic promotion 
into this state will require a (rather
large) energy of almost purely electronic origin.

In the  $Q=-2$ charge state,  the first antibonding state 
gets completely filled. In contrast to the $Q=-1$, 
 a large, markedly asymmetric relaxation occurs.
This can be seen in the structure model of Fig.
 \ref{fxmol}.  In this geometry 
atom Si$_1$ has moved  towards the vacancy 
by about 20 \% of the pristine unrelaxed bond length, whereas
 Si$_2$ has moved away from the vacancy  by about 3 \%. 
The tetrahedron centered on Si$_1$
remains essentially undistorted, while that centered on Si$_2$
is  distorded concurrently with the 
neighboring tetrahedron centered on a third Si atom, Si$_3$.
This effect is due to a large displacement of the bridging O atom.
 Due to the large relaxation energy gain and to the
associated electronic rehybridization (visible in Fig. 
\ref{f4rho-2}, where  the relevant charge density isosurfaces are 
displayed), the $Q=-2$ state is  lower in energy than the $Q=-1$
 (Fig.\ref{f1eform}) over most of the E$_{\rm F}$  range, despite the 
electron-electron repulsion. Thus a negative-U effect takes place.
The --2 state becomes favored over the neutral at a Fermi level of
5.5 eV.

At a  Fermi level about 0.2 eV higher the $Q=-3$ state charge
becomes favored. Si$_1$ moves  towards the vacancy site by an
additional 7 \%  (its distance from the vacancy  is  about
30 \% shorther than the unrelaxed bond length), while   Si$_2$  
moves closer by 2 \% (its distance from the vacancy  is  about
1 \% longer than the unrelaxed bond length).
The stability of the $Q=-3$ state over  the $Q=-2$ state for
a wide range of E$_{\rm F}$ is somewhat puzzling in view of 
the large repulsion to be expected among electrons in the center.
 The prominent feature of this state is indeed that the additional
electron minimizes the Coulomb repulsion with the preextant ones by
occupying a state in the interstice 
between the vacancy-adjacent Si$_1$ and Si$_2$ atoms, and 
the Si$_3$ atom belonging
to the tetrahedron neighboring to Si$_2$ and closest to
the vacancy. The distance between  Si$_2$
and Si$_3$ is reduced to only 4.3 a.u.,
to be compared with a distance of 5.8 a.u. in the
unrelaxed structure. This is due to the aforementioned
distortion of the Si$_2$-O-Si$_3$ bridge
linking the two tetrahedral units, whereby the  corresponding angle
decreases from the usual  144$^{\circ}$ to only about 80$^{\circ}$.
While  this distortion  is in fact already qualitatively 
present in the $Q=-2$ charge state,
 the additional electron in the $Q=-3$ state is effectively
bound by Si$_3$, as can be seen from Fig. \ref{f5rho-3}. 
In view of its electronic structure,  the --3 charge state may be
a candidate  E$'$-like center (see below).

The $Q=-3$ state is stable up to extreme $n$-type conditions
(Fermi level $\sim$ 8.1 eV),
whereby the $Q=-4$ state sets in. The fourth electron fills in the
state whose occupation was initiated with Q=--3, 	
with marginal additional relaxation. The charge density and geometry
are very close to the Q=--3. 

By the time one gets to  extreme $p$ or $n$-type
 conditions, the formation energies of the vacancy become 
very small. Vacancy-driven Compensation of acceptor and donor impurities
will be very efficient, and  we expect
 that $n$-type doping of  quartz will only be
achieved with extreme difficulty, if at all.
 
\subsection{Franck-Condon shifts}

The
Franck-Condon (FC) shift $\cal{F}$$_{Q,Q'}$ is defined as
the  total energy difference 
between the system in charge state $Q$ in its own equilibrium
configuration and the same system in the equilibrium configuration of
charge state $Q'$:  \begin{equation}{\cal{F}} _{Q,Q'} = E_{\rm tot}^Q
[{\bf R}_{Q'}]  - E_{\rm tot}^Q [{\bf R}_{Q}]\, . \end{equation}
The FC shift is symmetric upon
exchange of $Q$ and $Q'$ if the dependence of the total 
energy on configurational coordinates is quadratic and with identical force
constants in the two charge states. However, given the large distortions
 involved in the present case, the FC shift might be  asymmetric
(expecially for large shifts)
both  because of anharmonicity and of force-constant differences.
 These expectations are indeed verified
in our calculations. We find large asymmetries for large shifts,
 which are consistent with force constants differing by  factors of 
up to about $\sqrt{2}$ assuming a single quadratic 
configurational cooordinate. We express for convenience 
the FC  shifts between the non-positive (non-negative) charge states
as a matrix $\cal{F}^{\rm neg-states}$ ($\cal{F}^{\rm pos-states}$)
  whose elements
 $\cal{F}$$_{ij}^{\rm neg-states}$
 ($\cal{F}$$_{ij}^{\rm pos-states}$)
are the FC shifts for the $Q=-i$ ($Q=i$) state in the
geometry of the  $Q=-j$ ($Q=j$) state; this matrix
would be  symmetric in the symmetric-FC hypotesis,
and its diagonal elements are null. For negative states,
with  $Q$ between 0 and --3, we have
\[ \cal{F}^{\rm neg-states} = 
\left(
\begin{tabular}{cccc}
0 & 0.60 & 5.20 & 6.38 \\
0.37 & 0 & 1.21 & 1.93 \\
2.51 & 1.81 & 0 & 0.18 \\
3.17 & 2.20 & 0.18 & 0 \\
\end{tabular}
\right),
\]
while for positively charged states ($Q$ between  0 and +3) 
we have
\[ \cal{F}^{\rm pos-states} = 
\left(
\begin{tabular}{cccc}
 0    & 1.35   & 6.23  &   \\
0.64 & 0   & 1.76 &  \\
3.78    & 3.81 & 0 & 0.63 \\
4.41 	& 2.90 & 0.63 & 0 \\
\end{tabular}
\right),
\]
 all  energies being expressed in eV. Note that only the
 $\cal{F}$$_{i,i\pm1}$  elements are needed in the
discussion to follow.

\subsection{Optical transitions}

The optical  transition energy is the
total energy difference of the two (differently 
charged) initial and final 
systems in a fixed geometry.
Alternatively (but equivalently), optical levels can
be expressed as combinations of thermal levels and  FC shifts. 
We only consider one-electron (i.e. first-order) processes. This 
kind of process amounts to transferring only one electron at a time 
into, or out of, a defect level.

 The transitions 
we consider are  absorption (an electron is transferred 
from the valence states reservoir into the defect,
or transferred from the defect to the conduction band
via an internal excitation of e.g. the --3 state into the --2 state),
recombination (the electron drops
into the defect level from a reservoir of e.g. optically-excited
conduction electrons), and single-particle 
deexcitation of a metastable  charged center  by recombination 
of  an electron into the valence band.
We  assume that all processes involved, e.g. successive  absorption
transitions ($0\rightarrow -1$, $-1\rightarrow -2$,
$-2\rightarrow -3$ ...), take place on time scales shorter than those of 
lattice vibrations, so that (e.g.) absorption occurs in a frozen structure.
This assumption seems plausible in view of typical phonon frequencies
and optical transition times, especially in large excitation 
densities.  We therefore assume that the structure remain
  frozen in the equilibrium configuration of a given state $Q_f$
(typically the initial state) throughout the transition cascade. 

If the ground state of the defect is charge state $Q$,
 in terms of thermal levels $\epsilon[Q/Q']$
and FC shifts  $\cal{F}$$_{Q,Q'}$  the optical absorption  
energy for electronic transitions  from the valence  band
into charge state  $Q-1$ is
\begin{equation}
\delta_Q^{\rm abs} =  \epsilon[Q/Q-1]  
+ {\cal{F}} _{Q-1,Q_f}
- {\cal{F}} _{Q,Q_f}.\end{equation}
For internal absorption excitations (transforming e.g. $Q=-2$ into $Q=-1$
by electron excitation in the conduction band)
we have instead
\begin{equation}\delta_Q^{\rm int-abs} = E_{\rm gap} 
- \epsilon[Q-1/Q]  
+ {\cal{F}} _{Q,Q_f}
- {\cal{F}} _{Q-1,Q_f}. \end{equation}  
Within the same assumptions, for the conduction band-to-acceptor 
 recombination  we have {\it emissions} at
\begin{equation}
\delta_Q^{\rm rec} = 
E_{\rm gap} - \delta_Q^{\rm abs}
\end{equation} 
with 
$E_{\rm gap} = 8.9$ eV, the experimental gap \cite{add9-expgap} 
of $\alpha-$SiO$_2$.
Once again we implicitly assumed that subsequent  recombinations 
are  instantaneous on the scale of ionic relaxation.

Deexcitation is slightly more
complicated. Suppose that the vacancies have been  occupied 
with a number of electrons e.g. between 1 and 4. Assume further 
that the lifetimes of each charge states are larger than the 
 inverse of typical phonon frequencies ($\sim$ 10 THz). This 
assumption seems 
again admissible in view  of typical decay constants  of observed 
luminescence bands in silica \cite{decay} being of order 10$^{-5}$
to 10$^{-10}$ sec. Then,  each charge state can relax to its own 
ground state before deexcitation. For each vacancy in a given charge 
state, a cascade of deexcitation processes occurs, 
whereby each   charge state (between the initial one,
say $Q=3$, to $Q=0$)  is occupied in sequence, and
it loses one electron to the valence band while in {\it 
its own} equilibrium geometry. Expressing again total-energy 
differences in terms of thermal levels and FC shifts, we 
obtain  the dexcitation emission energies from charge state 
$Q$ as 
\begin{equation}\delta_Q^{\rm deex} = \epsilon[Q/Q-1] 
- {\cal{F}} _{Q,Q-1}.\end{equation}
For internal deexcitation the formula reads
\begin{equation}
\delta_Q^{\rm int-deex} = E_{\rm gap} - \epsilon[Q/Q-1] 
+ {\cal{F}} _{Q-1,Q}.\end{equation}
Here we did not consider selection rules possibly
preventing some of the transitions involved, and remain
 with an open question on this issue.

As mentioned above, quantitative predictions depend on 
which   ground state is actually realized. 
This makes the optical properties of this system
quite rich and complex.
In Table I we  summarize the main groups of
 absorption and emission transitions involving 
different ground states.
 If the Fermi level  
is at mid-gap (i.e. the solid is undoped), the  ground state is 
the neutral state. In moderate $n$-type conditions, the ground state
will typically be the --3 state.
The relevant  thermal levels are 
$\epsilon[0/-1]$ = 6.24 eV,
$\epsilon[-1/-2]$ = 5.47 eV, and
$\epsilon[-2/-3]$ = 5.82 eV. The 
FC shifts have been  given previously.
 In $p$-type conditions, the ground state
will be the +3 state.
The relevant  thermal levels are 
$\epsilon[+3/+2]$ = 2.48 eV,
$\epsilon[+2/+1]$ = 2.69 eV, and
$\epsilon[+1/0]$ = 0.72 eV (the 
FC shifts have been  given previously).

For the neutral state, the  absorption band is centered around
6.6 eV and the emission band at about 2.3 eV. 
For the --2 charge state there is an absorption band at about 6.3 eV,
while the emission one is centered around 3.0 eV. 
For the $Q=-3$ state, the absorption bands are 
located around 3.2 eV, 
6.0 eV,  and 7.1 eV. These correspond to internal excitations 
 $-3 \rightarrow -2$,
 $-2 \rightarrow -1$, and
 $-1 \rightarrow 0$ respectively.
While  calculated line intensities are not available,
we  speculate that the latter components may give rise to a
band centered  around 6.3 eV,  close to an experimentally observed 
absorption band (6.2 eV) usually assumed to be related to the E$'$
center. 

On the other hand, the emission bands are around  2.7 eV, close to the
2.7/2.8 eV emission band generally attributed to the oxygen vacancy,
and at 3.2 eV, where a band is also observed experimentally  in silica.
For the $Q=+3$ state, absorptions (either normal or internal) 
are predicted at 2.2 eV, 3.1 eV and 5.0 eV, 
while emissions are at 1.9 eV and 0.9 eV.

\subsection{ The $Q$=--3 as an E$'$-like center}

We note at this point that
the  triply-negative charge state of the oxygen vacancy 
exhibits several of the characteristic features of the
 E$'$ center \cite{add4-fowler,add10-weeks,add11-warren} 
in $\alpha$-quartz.
Both the transition generating  the center (if the
starting ground state is the neutral vacancy) and that characteristic
of the center itself are correctly predicted around  6 eV. While such
transition energies appear also in the predicted spectra of
other possible ground states,  only the electronic structure
of the $Q$=--3 state seems compatible with
 hyperfine-field and EPR measurements \cite{add10-weeks}. These
experiments indicate the presence of two weak hyperfine interactions
and a  
strong one associated with the oxygen vacancy. In
 our --3 state  the weak signals can be attributed to Si$_1$
and Si$_2$, while the strong signal comes from the unpaired  
electron prevailingly bound to Si$_3$. Such attributions could 
 not be made for the $Q$=+3, 0, and --2 ground states.
The  unpaired electron lobe is approximately directed along the
direction of the ``short bond'' \cite{add3-pasq,add10-weeks} in the ideal
structure, and as can be seen from Fig. \ref{fxmol}, Si$_3$ is close
to three fold coordinated, as it is generally accepted that
E$'$ should be  \cite{grisc}.

The main objection to identifying  the $Q$=--3 state of the vacancy
with E$'$  comes from  experiments showing that
E$'$ is positively charged in MOS Si-SiO$_2$ 
junctions \cite{add11-warren}. 
However, this result is not always reproduced  in other instances 
\cite{jap}. It would be interesting to analyze further these
experiments in the light of our results, taking into account for instance 
band offset effects at the Si/SiO$_2$ interface.
As to magnetic  data, recent calculations \cite{add3-pasq} of the
 hyperfine tensor  of a distorted variant of  the singly positive
 vacancy have resulted  in good agreement with the data pertaining to
the E$'$ center \cite{add10-weeks}, giving  support to
 the attribution of E$'$ to a positive 
charge state of the vacancy. On the other hand our results indicate
 that  the $Q$=+1 vacancy is unstable towards becoming triply charged,
and that the $Q$=+3 state appears to be a valence resonance, lacking
features which could explain  the peculiar magnetic signature of E$'$.

It is clear that  additional work is needed to  resolve this
 puzzle.  Nevertheless, the 3-- oxygen vacancy seems to carry
several E$'$-like features, and it may be 
considered as a possible new variety of E$'$ defect.

\section{Summary}

We have presented ab initio results on oxygen vacancies in $\alpha-$SiO$_2$.
The ground states for the vacancy as function of background
doping  are the triply-charged positive, neutral, and triply-charged
negative states.  The high-charge states are stabilized by large,
asymmetric distortions involving at least three tetrahedral units
neighboring to the vacancy.  In connection with the possible different
ground states we predict optical absorption energies in the 3.2 eV and
6-7 eV ranges, and emissions in the 2.3-3.2 eV range. Based on
predicted optical transitions, geometry, and  electronic structure, 
we  suggested a possible new  E$'$-like center, namely the
triply-charged negative oxygen vacancy. 

\section*{Acknowledgements}

We thank  David Vanderbilt for 
his  codes (coauthored in part by R.
D. King-Smith),  used in this work in a 
modified version. We thank Fabio Bernardini for useful
discussions. This work was supported in part by  CINECA Bologna
through Supercomputing  Grant 95-1506. C. M. C. was supported by the
Istituto Nazionale di Fisica  della Materia within Progetto Sud,
funded by the  European Union. Exchange of correspondence with
A. Pasquarello provided usefuls insights.

\begin{table}[h]
\begin{center}
\begin{tabular}{|c|c|c|}
Ground state &  absorptions & emissions \\
\hline
Q = +3 & 2.2 / 3.1  / 5.0 $\pm$ 0.2  & 1.9 / 0.9 $\pm$ 0.1\\
Q = 0 & 6.6 $\pm$ 0.2 & 2.3 $\pm$ 0.2 \\
Q = --2 & 6.3 $\pm$ 0.3 & 3.0 $\pm$ 0.1 \\
Q = --3 & 3.2 / 6.0 / 7.1 $\pm$ 0.2 & 2.7 $\pm$ 0.2 / 3.2 $\pm$ 0.2 \\
\end{tabular}
\caption{Summary of optical 
transitions (in eV) associated to the oxygen vacancy 
in $\alpha-$SiO$_2$}.
\end{center}
\end{table}

\begin{figure}
\caption{Formation energy of the oxygen vacancy in $\alpha$-SiO$_2$
as a function of the Fermi level.}
\label{f1eform}
\end{figure}

\begin{figure}
\caption{Charge density isosurfaces for the neutral state of the 
oxygen vacancy. The surface shown corresponds to a density value of
0.02 electrons/bohr$^3$.}
\label{f2rho0}
\end{figure}

\begin{figure}
\caption{Charge density isosurfaces for the singly-negative
charge state of the 
oxygen vacancy. The surface shown corresponds to a density value of
0.006 electrons/bohr$^3$.}
\label{f3rho-1}
\end{figure}

\begin{figure}
\caption{Relaxed geometry of the $Q=-2$ state of the oxygen vacancy
(black atoms and thick connecting lines).
The   geometry of the neutral vacancy is  given for reference
(white atoms and thin connecting lines).}
\label{fxmol}
\end{figure}

\begin{figure}
\caption{Charge density isosurfaces for the doubly-negative
charge state of the 
oxygen vacancy. The surface shown corresponds to a density value of
0.02 electrons/bohr$^3$.}
\label{f4rho-2}
\end{figure}

\begin{figure}
\caption{Charge density isosurfaces for the triply-negative
charge state of the 
oxygen vacancy. The surface shown corresponds to a density value of
0.0045 electrons/bohr$^3$.}
\label{f5rho-3}
\end{figure}


\begin{references}
\bibitem{jap}
See Refs. [52] to [56] of our Ref. \protect\cite{add11-warren}.
\end{references}
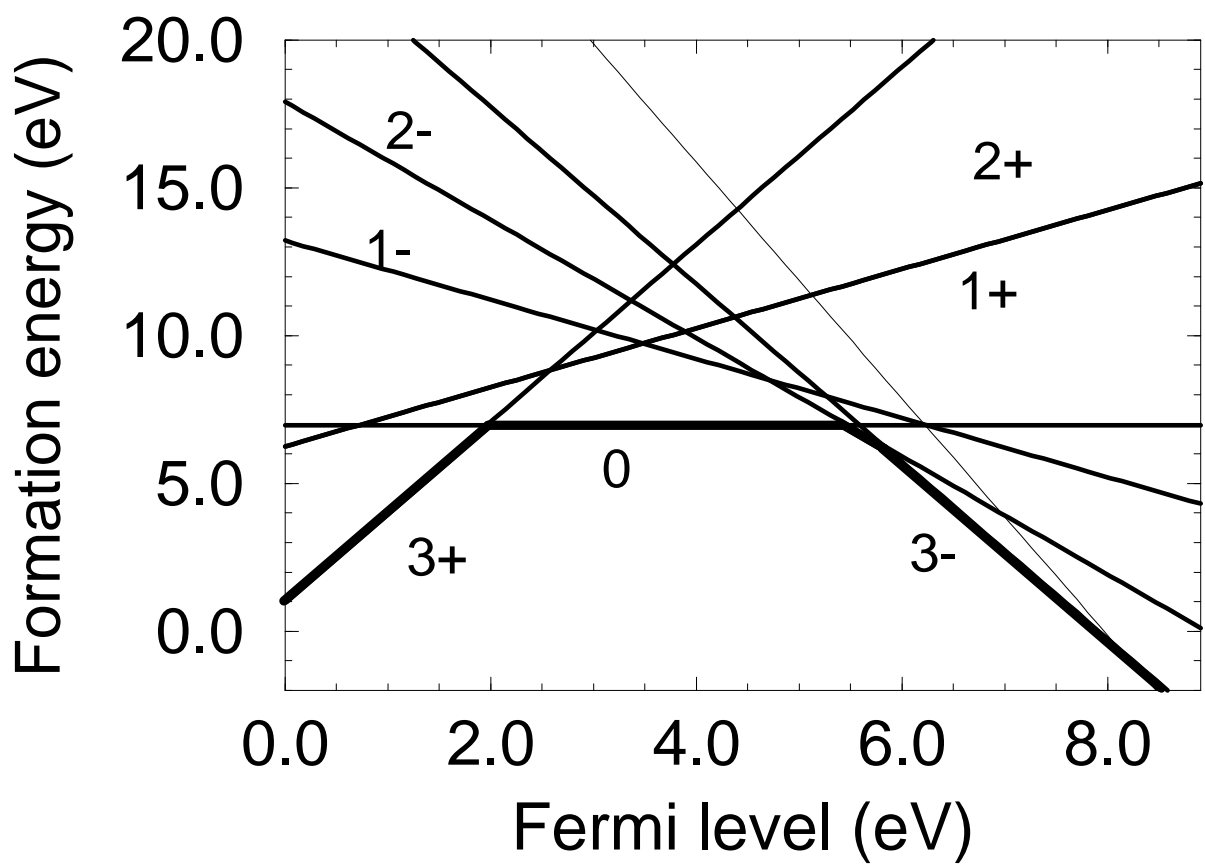
\end{document}